# Efficient Time-Domain Simulation of USV Motions in Short-Crested Irregular Waves Using an IRF-Based Framework

*Fei Duan*[a,b]*, Zihao Wang*[c*]*, Yaohua Zhou*[d]*, Qing Xiao*[b]
[a]State Key Laboratory of Ocean Engineering, Shanghai Jiao Tong University, Shanghai 200240, China;
[b] Department of Naval Architecture, Ocean and Marine Engineering, University of Strathclyde, Glasgow G4 0LZ, Scotland, United Kingdom;
[c] Institute of Artificial Intelligence, Shanghai University, Shanghai, 200444, China;
[d] Rules & Technology Center of China Classification Society, Shanghai 200240, China;

## ABSTRACT

Traditional time-domain prediction methods for vessel motions in irregular waves often require the superposition of responses from numerous regular-wave components, resulting in high computational cost for long-duration simulations and real-time applications. This challenge becomes more significant for unmanned surface vehicles (USVs), where efficient and realistic motion prediction is important for seakeeping assessment, simulation-based testing, and intelligent control algorithm development. To address this issue, an efficient IRF-based time-domain framework is applied for predicting vessel motions in short-crested irregular waves. Hydrodynamic loads, including Froude–Krylov, diffraction, and radiation forces, are obtained from frequency-domain analysis and transformed into the time domain. The resulting framework directly evaluates instantaneous motion responses in irregular waves through convolution-based force reconstruction, thereby significantly improving computational efficiency. Weak nonlinear restoring effects are incorporated through instantaneous wetted-surface pressure integration, while directional wave spectra are introduced to simulate realistic short-crested sea conditions. The framework is first validated against experimental measurements of an offshore supply vessel (OSV) in long-crested beam irregular waves, demonstrating good agreement in transient roll responses and motion statistics. Further validation is performed using full-scale measurements of a USV operating in real sea conditions. Predicted significant amplitudes, mean zero-crossing periods, standard deviations, and motion time histories show good agreement with the measured results. In addition, the influence of directional-spectrum discretization on motion prediction is investigated. The results indicate that motion amplitudes exhibit moderate sensitivity to directional resolution, whereas motion periods remain relatively insensitive. A directional discretization interval of 30 deg is found to provide a reasonable balance between prediction accuracy and computational efficiency. The present framework provides a computationally efficient approach for high-fidelity time-domain prediction of vessel motions in realistic directional irregular-wave environments, with potential applications in real-time motion prediction and simulation-based evaluation of USVs.

## 1. Introduction

Realistic prediction of transient vessel motions in directional irregular seas remains a major challenge for high-fidelity maritime simulation environments, particularly for unmanned surface vehicles (USVs) that are highly sensitive to wave disturbances. With the rapid development of autonomous maritime systems, high-fidelity simulation frameworks have become increasingly important for system testing, operational assessment, and intelligent control algorithm development. In realistic sea conditions, coupled heave, pitch, and roll motions of USVs may exhibit strong transient characteristics under multidirectional wave excitations, which can significantly influence navigation safety, sensor stability, and mission performance. Therefore, efficient and physically realistic prediction of instantaneous USV motions in directional irregular seas is essential for constructing reliable simulation environments and supporting the development of autonomous maritime systems.

Time-domain prediction of ship motions is commonly performed using either potential-flow-based methods or fully nonlinear CFD approaches. Among potential-flow-based approaches, the impulse response function (IRF) method has been widely used to transform frequency-domain hydrodynamic coefficients into the time domain for predicting nonlinear and transient ship motions [1][2]. Existing studies have demonstrated the capability of IRF-based methods for simulating ship responses in regular waves [3]. Begovic et al. [4] and Spyrou et al. [5] reconstructed instantaneous ship motions in long-crested irregular waves through the superposition of responses induced by multiple regular wave components with different amplitudes, frequencies, and phases. However, for directional irregular waves, the computational cost increases substantially due to the large number of directional wave components required in the wave spectrum discretization. As a result, efficient prediction of transient ship motions in short-crested irregular seas remains computationally challenging for long-duration simulations and real-time applications.

Spectral analysis methods provide an alternative approach for predicting ship motions in irregular waves [6][7]. Jiao et al. [8] and Tang et al. [9] combined motion response amplitude operators (RAOs) obtained from linear potential flow theory with directional wave spectra to estimate ship motion statistics under short-crested irregular waves. Although these approaches are computationally efficient for predicting motion statistics, they are generally unsuitable for simulating transient instantaneous motions under time-varying or nonlinear wave excitations, since nonlinear hydrodynamic effects associated with instantaneous wetted-surface variations cannot be captured directly. In recent years, machine-learning-based approaches have also been applied to ship and USV motion prediction. Mounet et al. [10][11] estimated USV motions using onboard inertial measurements, while Zhang et al. [12], Luo et al. [13], and Lee et al. [14] developed neural-network-based models for predicting nonlinear ship motions in irregular waves. Although these methods achieved reasonable prediction accuracy under specific operating conditions, their performance generally depends strongly on training datasets and may have limited applicability outside trained conditions.

Fully nonlinear CFD approaches provide another important framework for predicting transient hydrodynamic responses in directional irregular waves. Cao and Wan [15], Kumar et al. [16], Huang et al. [17] and Hong et al. [18] constructed CFD numerical wave basins for short-crested or crossing-wave conditions and investigated nonlinear wave–body interaction phenomena under multidirectional wave excitations. Zhang et al. [19] further analyzed the speed loss and propulsion performance of a trimaran in short-crested head seas, while Ma et al. [20] simulated maneuvering motions of ships in irregular wave environments. Although CFD methods can provide high-fidelity predictions of nonlinear hydrodynamic responses, accurate simulations of directional irregular waves generally require fine free-surface meshes and small time steps, particularly for short-wavelength and high-frequency wave components [21][22]. As a result, CFD-based simulations remain computationally expensive for long-duration simulations and real-time high-fidelity simulation environments.

The above studies indicate that existing approaches generally involve a trade-off between computational efficiency and physical fidelity. Spectral and data-driven approaches are computationally efficient but have limited capability in directly predicting transient nonlinear motions, while CFD and conventional irregular-wave superposition approaches can provide more realistic transient responses at significantly higher computational cost. Therefore, there remains a lack of computationally efficient and physically consistent approaches capable of directly predicting instantaneous ship motions in directional irregular seas for high-fidelity simulation applications.

To address these challenges, this study presents an efficient IRF-based time-domain framework for predicting instantaneous USV motions in short-crested irregular seas. Long-crested irregular wave time histories are generated from discretized wave spectra, while hydrodynamic loads, including Froude–Krylov, diffraction, and radiation forces, are transformed from the frequency domain into the time domain through IRFs. Unlike conventional approaches that reconstruct irregular-wave responses through the superposition of numerous regular-wave simulations, the present framework enables efficient prediction of instantaneous ship motions in irregular waves while improving computational efficiency. Weak nonlinear restoring effects are incorporated through instantaneous wetted-surface pressure integration using a B-spline surface representation. Directional wave spectra are further introduced to simulate realistic short-crested sea conditions. The framework is validated against experimental measurements of an offshore support vessel (OSV) in long-crested beam irregular waves and full-scale measurements of a USV operating in real sea conditions. In addition, the influence of directional spectrum discretization on motion prediction accuracy and computational efficiency is investigated.

The above studies indicate that existing approaches generally involve a trade-off between computational efficiency and physical fidelity. Spectral and data-driven approaches are computationally efficient but are generally limited in reproducing transient nonlinear motion responses, whereas CFD-based methods and conventional irregular-wave superposition approaches can provide more realistic transient responses at significantly higher computational cost. Therefore, there remains a lack of computationally efficient and physically consistent approaches capable of directly predicting instantaneous vessel motions in directional irregular seas for high-fidelity simulation and real-time prediction applications.

In the present study, an IRF-based time-domain framework is applied to predict instantaneous USV motions in short-crested irregular waves. Long-crested irregular-wave time histories are generated from discretized wave spectra, while hydrodynamic loads, including Froude–Krylov, diffraction, and radiation forces, are transformed from the frequency domain into the time domain through IRFs. Unlike conventional approaches based on the superposition of numerous regular-wave responses, the present framework directly evaluates instantaneous motion responses in irregular waves through time-domain convolution, substantially reducing the computational cost of long-duration simulations. Weak nonlinear restoring effects are incorporated through instantaneous wetted-surface pressure integration using a B-spline surface representation. Directional wave spectra are further introduced to simulate realistic short-crested sea conditions. The framework is validated against experimental measurements of an OSV in long-crested beam irregular waves and full-scale measurements of a USV operating in real sea conditions. In addition, the influence of directional-spectrum discretization on motion prediction accuracy and computational efficiency is investigated.

## 2. Numerical framework

### 2.1 Time-domain equations of motion

To predict instantaneous USV motions in irregular waves while accounting for weak nonlinear restoring effects associated with instantaneous attitude variations, a time-domain framework based on the IRF method is employed in the present study. Frequency-domain hydrodynamic coefficients are first obtained and subsequently transformed into time-domain memory functions for evaluating radiation and wave excitation forces. The coupled ship motions are governed by the following 5-DOF time-domain

equations:

$$\sum_{j=2}^{6}\left[\left(m_{ij}+A_{ij}(\infty)\right)\dot{v}_j(t)\right]+\left(B_1+B_2\left|v_i\right|\right)v_i\delta_{i4}=F_i^{res}(t)+F_i^{FK+dif}(t)-F_i^{rad}(t)\;\; i=2,3\cdots6 \tag{1}$$

where $j$=2 to 6 correspond to the sway, heave, roll, pitch and yaw motions, respectively. $v_i$ is the instantaneous motion velocity. $m_{ij}$ is the mass and inertia matrix, and $A_{ij}(\infty)$ denotes the added mass at infinite frequency. $B_1$ and $B_2$ are the linear and nonlinear viscous damping coefficients associated with roll motion, respectively. $\delta_{ij}$ is the Kronecker delta. $F_i^{res}(t)$, $F_i^{FK+dif}(t)$ and $F_i^{rad}(t)$ denote the restoring, wave excitation and radiation forces, respectively.

To account for weak nonlinear restoring effects caused by instantaneous wetted-surface variations, the USV hull surface is represented using a Non-Uniform Rational B-Spline (NURBS) model generated in NAPA. Based on the instantaneous ship motions, the coordinates and normal vectors of the wetted surface panels are transformed into the global coordinate system. The instantaneous restoring force is then evaluated through pressure integration over the submerged hull surface:

$$\begin{cases} F_2^{res}(t)=-\sum_{i=1}^{N}P\left(r_i(t)\right)A_i n_i(t) \\ F_3^{res}(t)=-\sum_{i=1}^{N}P\left(r_i(t)\right)A_i n_i(t)-mg\cos\theta(t)\cos\phi(t) \\ F_j^{res}(t)=-\sum_{i=1}^{N}P\left(r_i(t)\right)A_i n_i(t)\times\left(r_G-r_i(t)\right)(j=4,5,6) \end{cases} \tag{2}$$

where $N$ is the total number of surface panels. $A_i$ is the area of the $i^{\text{th}}$ panel. $\boldsymbol{r}_i(t)$ and $\boldsymbol{n}_i(t)$ denote the centroid coordinate and unit normal vector of the corresponding panel in the global coordinate system, respectively. $P(\boldsymbol{r}_i(t))$ is the instantaneous pressure acting on the hull surface.

The radiation and wave excitation forces are evaluated in the time domain using the IRF method [2]. The corresponding formulations are expressed as:

$$F_i^{rad}(t)=\sum_{j=2}^{6}\int_0^t R_{ij}(t-\tau)v_j(\tau)d\tau \tag{3}$$

$$F_i^{FK+dif}(t)=\int_{-\infty}^{t}Q_i(\tau)\zeta(t-\tau)d\tau \tag{4}$$

where $\zeta(t)$ is the instantaneous wave elevation. $R_{ij}(t)$ and $Q_i(t)$ are the memory effect functions associated with the radiation and wave excitation forces, respectively.

### 2.2 IRF-based evaluation of wave excitation forces in irregular waves

In conventional time-domain simulations of weakly nonlinear ship motions in irregular waves, the instantaneous wave elevation is commonly decomposed into a large number of regular wave components using Fourier-based methods. The nonlinear Froude–Krylov (F–K) force associated with each wave component is then evaluated through instantaneous wetted-surface integration, and the total excitation force is reconstructed by superposing the contributions from all wave components. Yu et al. [23], Spanos et al. [24], Matusiak [25], Ma et al. [26], and Riesner et al. [27] adopted similar strategies for evaluating nonlinear wave excitation forces in time-domain ship motion simulations. Although these approaches can accurately capture nonlinear excitation effects, repeated wetted-surface integrations for numerous wave components lead to substantial computational cost, particularly for long-duration simulations in irregular seas.

To improve computational efficiency, the present study evaluates wave excitation forces directly in the time domain using an IRF-based formulation. Frequency-domain F–K forces and diffraction forces are transformed into time-domain convolution kernels through inverse Fourier transforms. The resulting framework enables direct evaluation of instantaneous wave excitation forces under irregular-wave excitations without reconstructing ship responses from numerous regular-wave simulations.

The time-domain convolution kernel of the wave excitation force for long-crested irregular waves is obtained through the inverse Fourier transform as follows:

$$Q_i(t) = \frac{1}{\pi}\int_0^{\infty}\Big[\big(d_i^R(\omega) + f_i^R(\omega)\big)\cos(\omega t) - \big(d_i^I(\omega) + f_i^I(\omega)\big)\sin(\omega t)\Big]d\omega \tag{5}$$

where $d_i^R$ and $d_i^I$ represent the real and imaginary parts of the wave diffraction force, respectively, while $f_i^R$ and $f_i^I$ denote the real and imaginary parts of the F-K force derived from the incident wave potential.

Because the convolution integration in Eq. (4) formally extends from $-\infty$ to the current time $t$, the memory effect is truncated to a finite discrete time window in the numerical implementation. After temporal discretization, the convolution integration is evaluated over a finite number of previous time steps, expressed as $N_m$. In the present study, $N_m$=9999 is adopted to ensure sufficient decay of the memory function before truncation, thereby preventing non-physical transient responses at the initial stage of the simulation. Accordingly, an initial pre-simulation time window is introduced to account for the fluid memory effect associated with previous wave states.

The instantaneous wave elevation may be obtained either from direct wave measurements or reconstructed from a discretized wave spectrum. When measured wave elevations are available, the instantaneous surface elevation at the $n^{\text{th}}$ time step is directly stored as $\zeta^{LC}\left(N_m + \frac{T}{\Delta t} - n\right)$, where $n$=0 to $n = 0,1\ldots,\left(N_m + \frac{T}{\Delta t}\right)$ and $T$ denotes the total simulation duration. Alternatively, when the wave elevation is reconstructed from the wave spectrum $S_\zeta(\omega)$, the corresponding discretized time history is expressed as:

$$\begin{cases} \zeta_{ai}^{LC} = \sqrt{2S_\zeta(\omega_i)\Delta\omega} & i = 1 \sim N \\ \zeta^{LC}(n) = \sum_{i=1}^{N}\zeta_{ai}^{LC}\cos\left(\omega_{ei}\left(N_m + \frac{T}{\Delta t} - n\right)\Delta t + \varepsilon_i\right) & n = 0 \sim N_m + \frac{T}{\Delta t} \end{cases} \tag{6}$$

Based on the convolution formulation, the discretized wave excitation force at time $t$ can be written as:

$$F_i^{FK+dif}(t) = \sum_{n=-N_m}^{t/\Delta t} Q_i(n)\zeta^{LC}\left(n + N_m + \frac{T}{\Delta t} - \frac{t}{\Delta t}\right)\Delta t \qquad t = 0 \sim T \tag{7}$$

In conventional irregular-wave simulations, the wave elevation time history is typically decomposed into numerous regular wave components using Fourier-based methods, and the corresponding wave excitation forces are evaluated separately for individual wave components. As illustrated in Fig.1, the present IRF-based framework avoids repeated wetted-surface integrations for individual regular wave components by directly evaluating the instantaneous wave excitation force through convolution integration of the irregular-wave elevation time history. This substantially reduces the computational cost associated with long-duration time-domain simulations in irregular seas while retaining the physical consistency of the motion response.

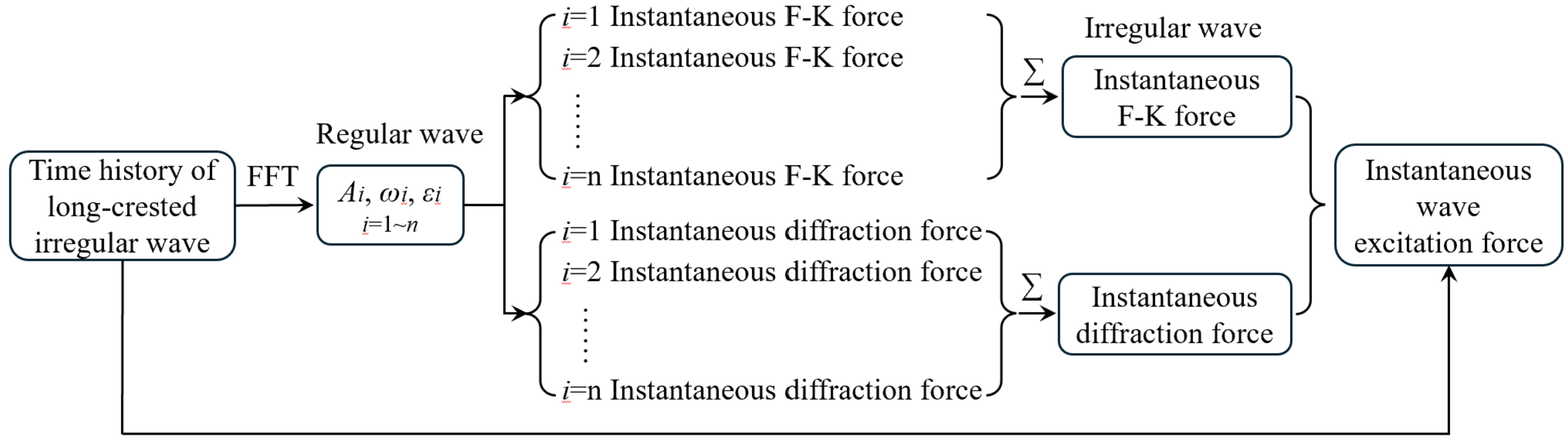

Figure 1 Comparison between conventional and present approaches for evaluating wave excitation forces in long-crested irregular waves

### 2.3 Directional modeling of short-crested irregular waves

To represent realistic sea conditions, ocean waves are modeled as a stationary random process described by a directional wave spectrum $R(\omega,\beta)$ [28]. The directional spectrum is constructed by combining a wave frequency spectrum $S_\zeta(\omega)$ with a directional spreading function $D(\beta-\beta_{\mathrm{m}})$:

$$R(\omega,\chi)=S_\zeta(\omega)D(\beta-\beta_{\mathrm{m}}) \tag{8}$$

$$D(\beta-\beta_{\mathrm{m}})=\begin{cases}\dfrac{2^{2s}}{\pi}\dfrac{\Gamma^2(s+1)}{\Gamma(2s+1)}\cos^{2s}(\beta-\beta_{\mathrm{m}}) & \beta_{\mathrm{m}}-\dfrac{\pi}{2}\le\beta\le\beta_{\mathrm{m}}+\dfrac{\pi}{2}\\ 0 & for(\text{others})\end{cases} \tag{9}$$

where $\Gamma$ is the Gamma function. $s$ is the directional spreading parameter. $\beta$ denotes the encountered wave direction of the long-crested irregular wave and $\beta_{\mathrm{m}}$ represents the encountered wave direction of the short-crested irregular wave. The influence of the spreading parameter on the directional spreading function is illustrated in Fig.2. In the present study, $s$=1 is adopted following the recommendation of IMO[29].

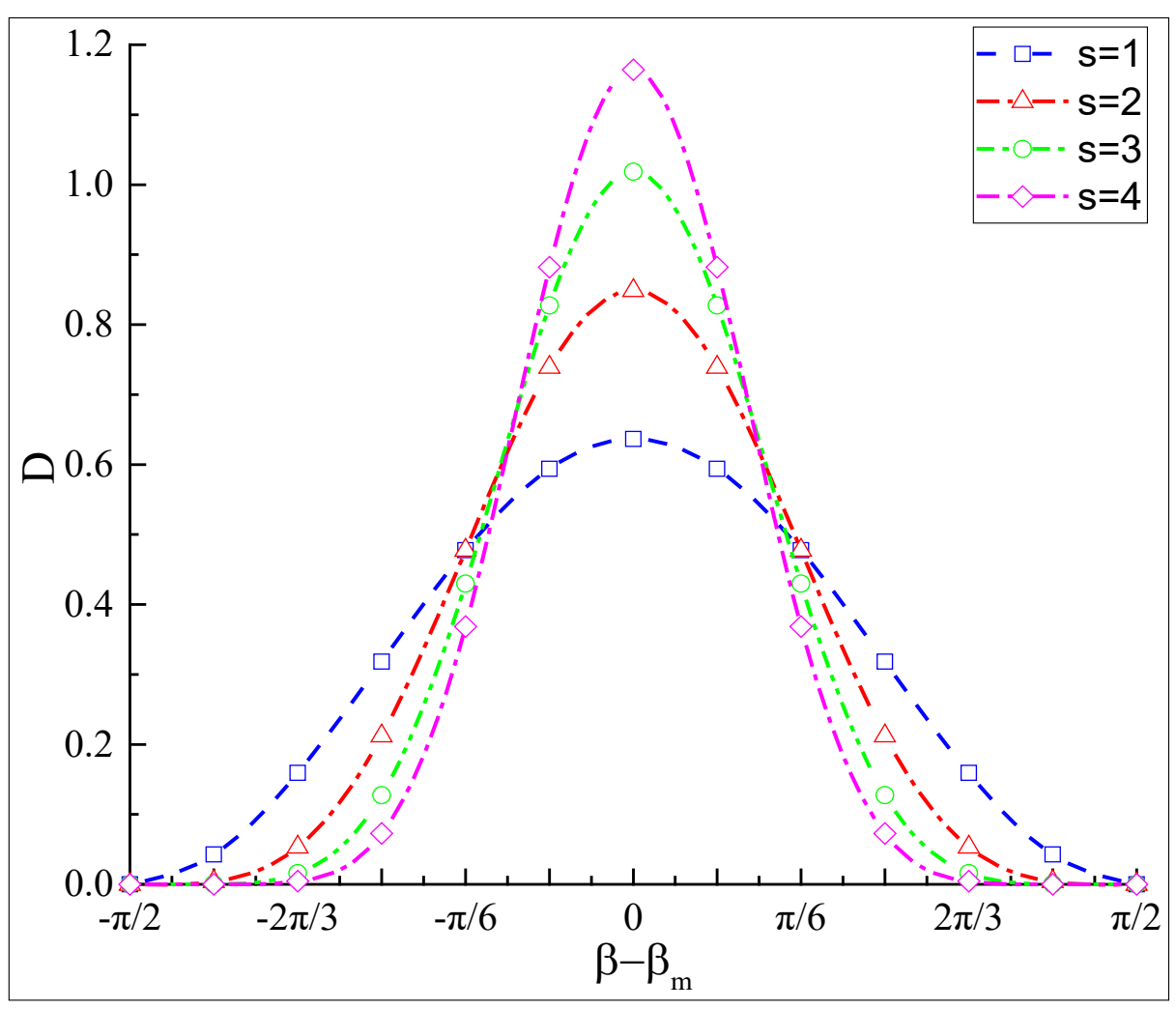

Figure 2 Directional spreading functions corresponding to different spreading parameters

After introducing the directional spectrum, the amplitude of the $(i, j)^{\text{th}}$ wave component in the short-crested irregular wave system can be expressed as:

$$\zeta_{aij}^{SC} = \sqrt{2S_{\zeta}(\omega_i)D(\beta_j - \beta_{\text{m}})\Delta\omega\Delta\beta} = \zeta_{ai}^{LC}\sqrt{D(\beta_j - \beta_{\text{m}})\Delta\beta} \qquad i = 1 \sim N, j = 1 \sim M \tag{10}$$

where $N$ and $M$ denote the numbers of discretized frequency and directional components, respectively.

Since the instantaneous elevation of a long-crested irregular wave can be represented by Eq. (6), the instantaneous surface elevation of a short-crested irregular wave is reconstructed through the superposition of long-crested irregular waves propagating in different directions:

$$\zeta^{SC}(t) = \zeta_j^{LC}(t, \beta_j) \sum_{\beta_j = -\pi/2}^{\pi/2} \sqrt{D(\beta_j - \beta_{\text{m}})\Delta\beta} \qquad j = 1 \sim M \tag{11}$$

where $\zeta^{SC}(t)$ is the instantaneous surface elevation of the short-crested irregular wave, while $\zeta_j^{LC}(t, \beta_j)$ denotes the instantaneous elevation of the long-crested irregular wave propagating in the $\beta_j$ direction.

In the present study, the wave directions are discretized over the range from −90 deg to +90 deg at intervals of 15 deg. Since the directional spreading function $D(\beta - \beta_{\text{m}})$ becomes zero at −90 deg and +90 deg, a total of 11 directional wave components are considered. When the discretization interval is increased to 30 deg, only 5 directional components are required, substantially reducing the computational cost associated with directional-wave simulations. Considering that the number of discretized wave directions remains relatively limited, the instantaneous ship motions in short-crested irregular waves are reconstructed through directional superposition of the corresponding motion responses in long-crested irregular waves:

$$\xi_i^{SC}(t) = \xi_j^{LC}(t, \beta_j) \sum_{\beta_j = -\pi/2}^{\pi/2} \sqrt{D(\beta_j - \beta_{\text{m}})\Delta\beta} \qquad i = 2, 3 \cdots 6, \;\; j = 1 \sim M \tag{12}$$

where $\xi_i^{SC}(t)$ denotes the instantaneous motion responses induced by short-crested irregular waves, with $j$=2 to 6 corresponding to the sway, heave, roll, pitch, and yaw motions, respectively, while $\xi_j^{LC}(t, \beta_j)$ represents the instantaneous motion responses associated with the $\beta_j$ long-crested wave direction.

Combined with the convolution-based evaluation of irregular-wave excitation forces described in Section 2.2, the directional superposition strategy provides an efficient approach for approximating transient ship motions in short-crested irregular seas while maintaining the primary directional characteristics of the wave-induced responses.

## 3. Motion validation in long-crested irregular waves

The accurate evaluation of wave excitation forces is essential for reliable prediction of ship motions in irregular waves. To assess the applicability of the present IRF-based framework, the instantaneous wave elevations measured in model tests are directly used as input to Eq. (7), ensuring that the numerical simulations are performed under the same wave realizations as those in the experiments. The predicted ship motion responses are then compared with the corresponding model-test measurements to evaluate the capability of the present approach in reproducing instantaneous motion responses in long-crested irregular waves.

### 3.1 Test vessel and wave conditions

The validation study is performed using an OSV, as shown in Fig. 3. The model scale ratio is 1:42, and the main dimensions are listed in Table 1.

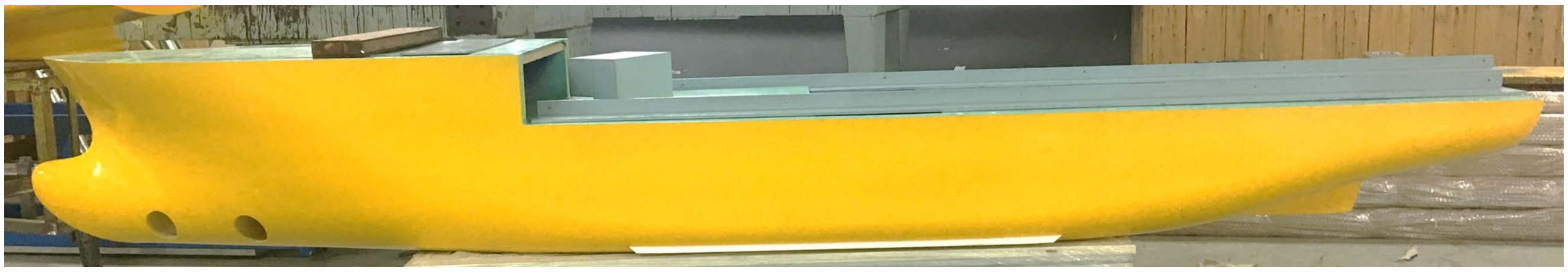

Figure 3 OSV model used in the towing-tank experiments

Table 1 Particulars of the offshore support vessel

| Particular | | Unit | Full scale | Model scale |
|---|---|---|---|---|
| Length between perpendiculars | $L_{PP}$ | m | 126.1 | 3.00 |
| Breadth | $B$ | m | 27.5 | 0.6545 |
| Height of main deck | $D$ | m | 11.88 | 0.2827 |
| Mean draught | $\overline{d}$ | m | 7.297 | 0.1737 |
| Trim angle | $\theta$ | deg | 0.166 | 0.166 |
| Mass | $m$ | kg | 18518017.8 | 243.85 |
| Height of center of gravity | $Z_g$ | m | 10.42 | 0.2481 |
| Roll natural period | $T$ | s | 12.28 | 1.895 |
| Roll moment of inertia | $I_{xx}$ | kg·m$^2$ | 2340255219 | 17.47 |
| Metacentric height | $GM$ | m | 3.74 | 0.089 |
| Water density | $\rho$ | kg/m$^3$ | 1025 | 1000 |

The model tests were conducted in the Multiple Function Towing Tank at Shanghai Jiao Tong University. Soft mooring lines were used to restrain the low-frequency drift motion of the vessel in the horizontal plane. A wave probe was installed at a transverse distance of 6.3 m from the vessel to record the instantaneous wave elevation. The experimental arrangement is shown in Fig.4.

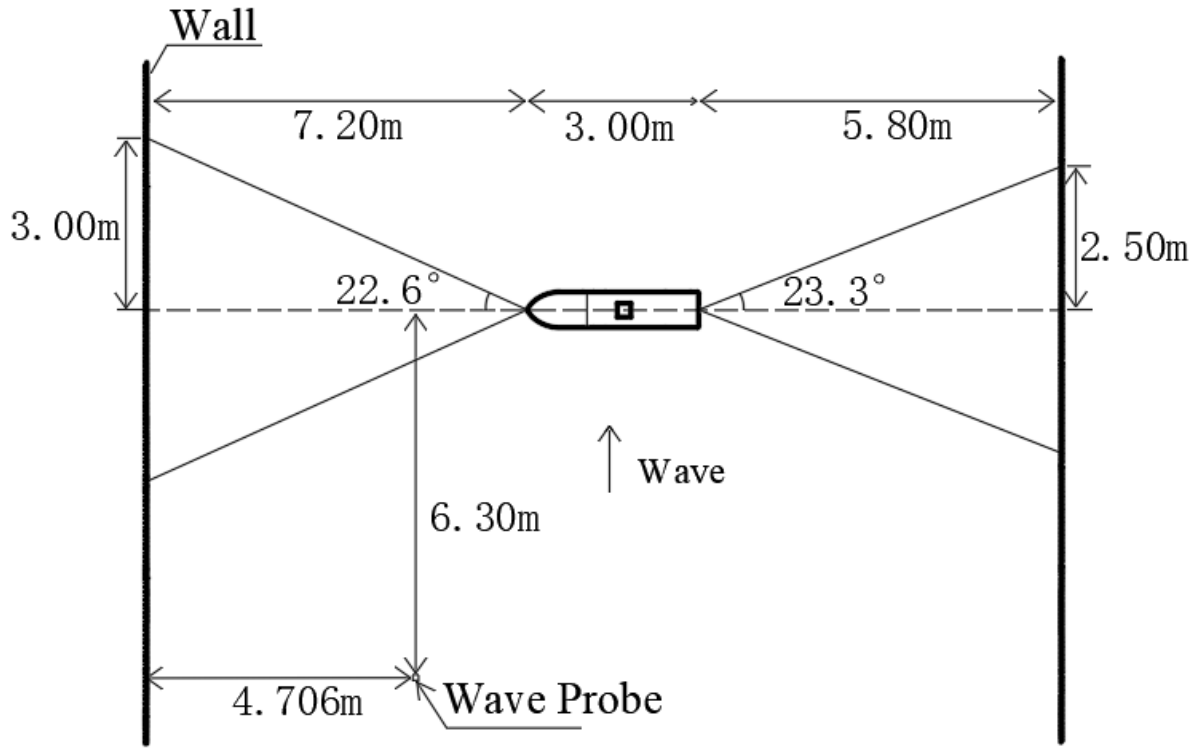

Figure 4 Experimental arrangement of the soft-mooring system

Table 2 summarizes the long-crested irregular-wave conditions used in the experiments. The irregular waves were generated using the ITTC two-parameter spectrum. Considering the roll natural period of the OSV, two average zero-crossing periods, $T_z$=11.5 s and $T_z$=12.5 s, were selected. The significant wave heights were set to 4.5 m, 5.5 m, and 6.5 m, corresponding to representative sea states in the North Atlantic wave scatter diagram.

Table 2 Wave conditions of Long-crested irregular wave

| | Full scale | | Model scale | |
|---|---|---|---|---|
| | Significant wave height $H_s$ (m) | Average zero-crossing period $T_z$ (s) | Significant wave height $H_s$ (m) | Average zero-crossing period $T_z$ (s) |
| Case 1 | 4.5 | 11.5 | 0.1178 | 1.8607 |
| Case 2 | 4.5 | 12.5 | 0.1178 | 2.0225 |
| Case 3 | 5.5 | 11.5 | 0.1440 | 1.8607 |
| Case 4 | 5.5 | 12.5 | 0.1440 | 2.0225 |
| Case 5 | 6.5 | 11.5 | 0.1702 | 1.8607 |
| Case 6 | 6.5 | 12.5 | 0.1702 | 2.0225 |

### 3.2 Validation of roll motions in long-crested irregular waves

Due to the shallow draft, open moonpool, and bow-thruster tunnels of the OSV, viscous and local flow-separation effects become significant, particularly in short-wave conditions. Consequently, wave excitation forces predicted solely using potential-flow theory may exhibit noticeable discrepancies at higher frequencies. To improve the accuracy of the first-order wave excitation forces, the potential–viscous hybrid approach developed by Duan et al.[30] is adopted in the present study. In this approach, CFD simulations are first performed for the fixed hull under a series of regular-wave conditions. The resulting CFD-based excitation forces are then used to correct the corresponding frequency-domain hydrodynamic coefficients obtained from potential-flow analysis. The corrected wave excitation forces per unit wave amplitude are presented in Fig.5.

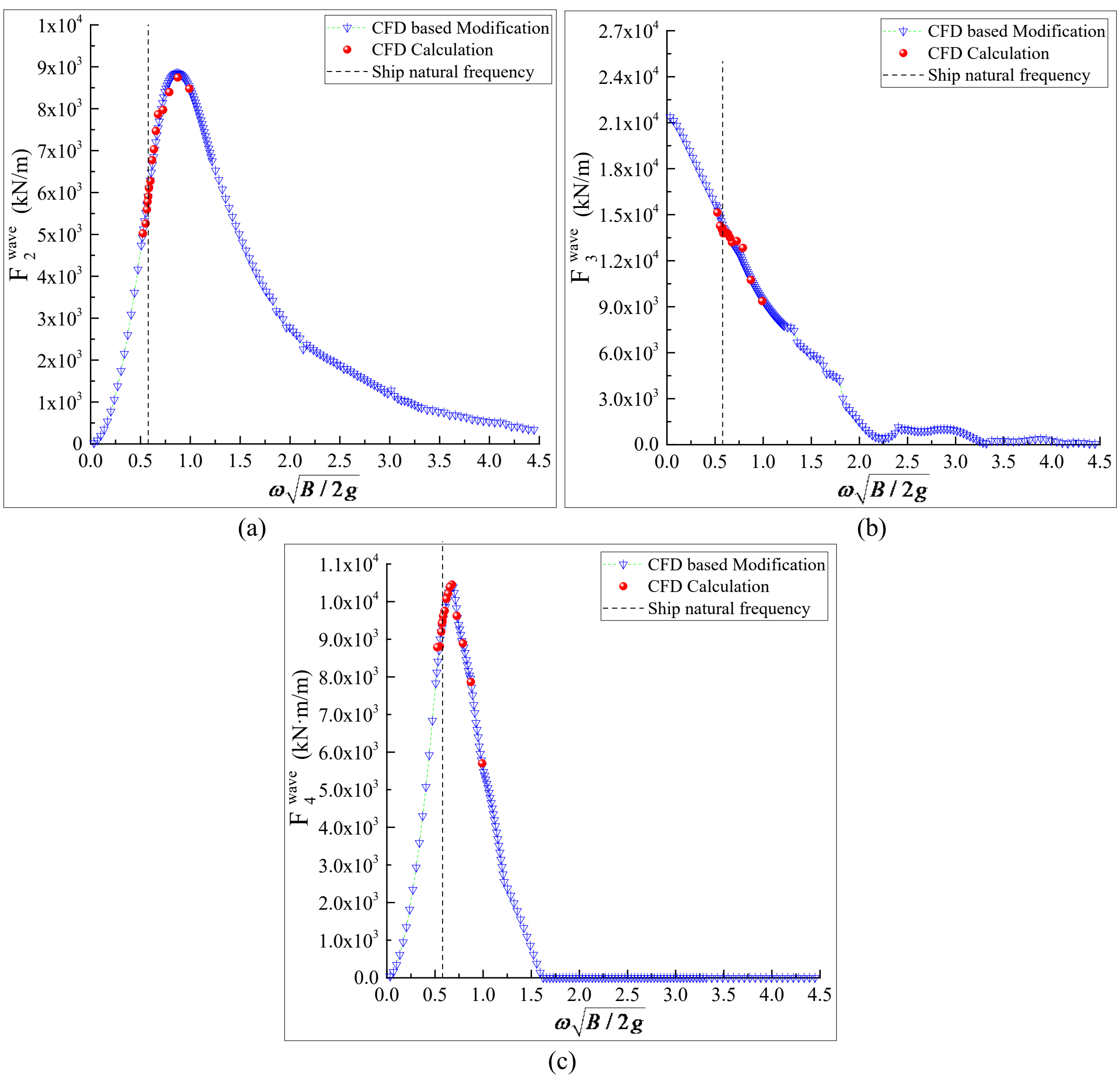

Figure 5 Corrected frequency-domain wave excitation forces per unit wave amplitude in surge (a), heave (b), and roll (c)

Among the vessel motions considered, roll motion is selected for detailed time-history comparison because it exhibits the strongest nonlinear characteristics and is particularly sensitive to irregular-wave excitation-force evaluation under beam-wave conditions. Fig.6 compares the simulated and measured roll-motion time histories in long-crested beam irregular waves. Under identical wave realizations, the numerical results show good agreement with the experimental measurements in terms of both instantaneous roll amplitudes and motion periods. The present IRF-based framework reproduces the dominant transient features of the roll response.

For most wave conditions, the differences between the predicted and measured maximum and minimum roll amplitudes remain within 5%, while the standard deviation error of the roll motion is generally below 2%. These results demonstrate the capability of the present framework in reproducing transient roll responses of the OSV under long-crested irregular-wave conditions.

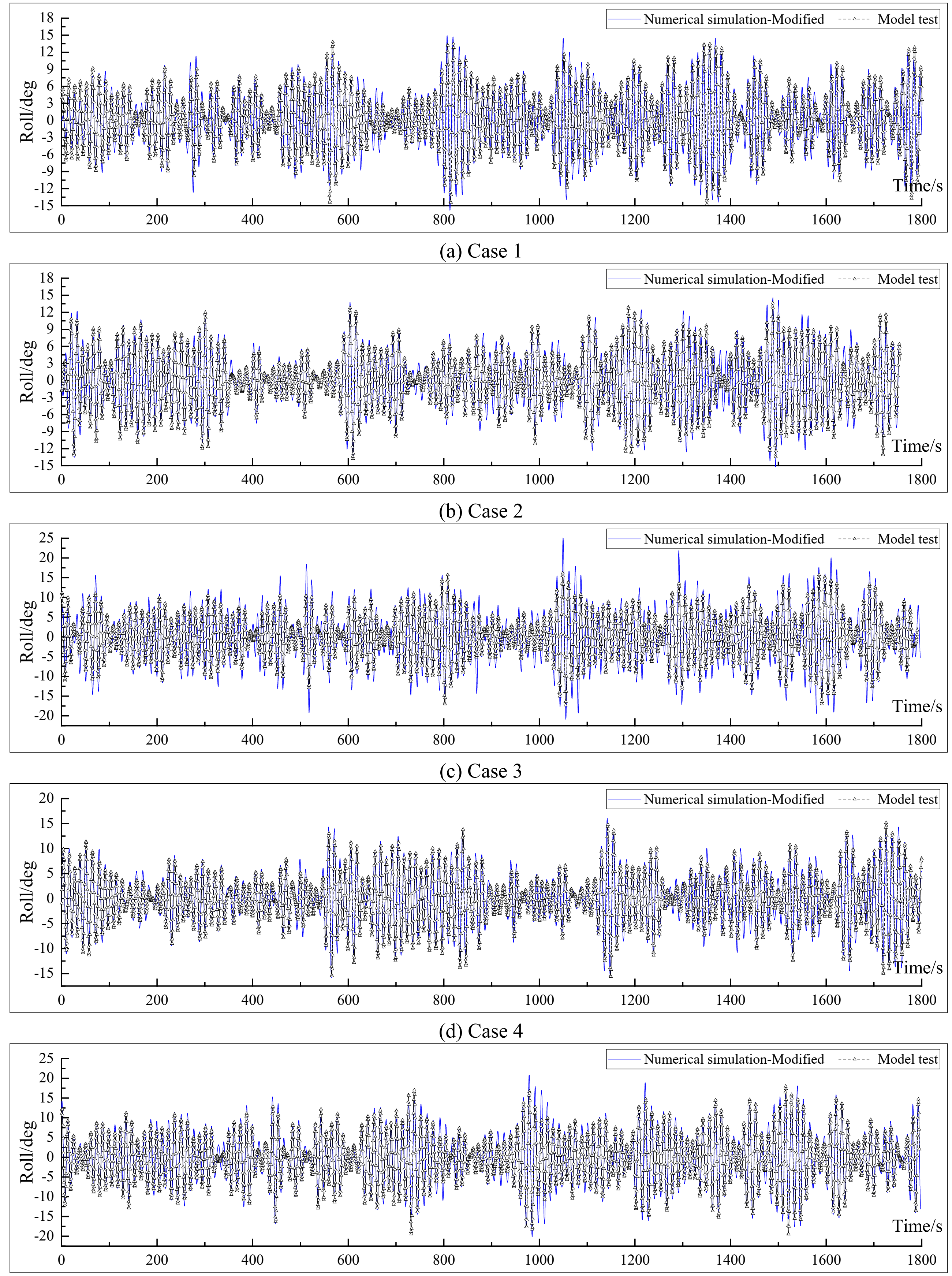

(a) Case 1

(b) Case 2

(c) Case 3

(d) Case 4

(e) Case 5

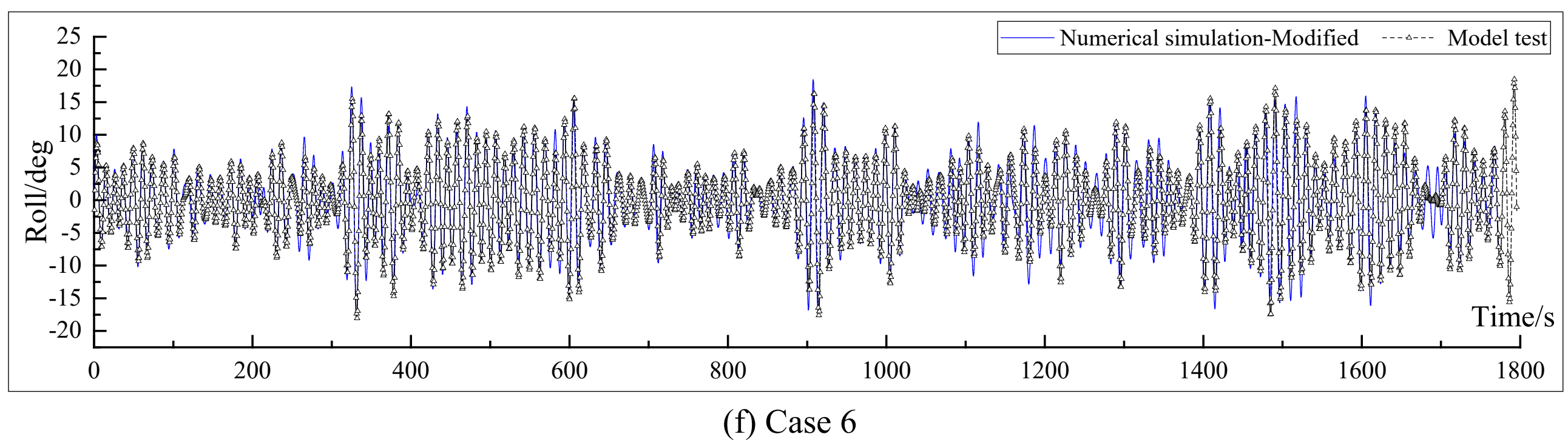

(f) Case 6

Figure 6 Comparison of simulated and measured roll motions of the OSV in long-crested beam irregular waves

### 3.3 Computational performance

Table 3 summarizes the computational performance of the present framework. The total duration of the simulated motion responses under the considered wave conditions is approximately 1800 s, corresponding to 30 min of physical simulation time. All computations were performed on a single CPU core (Intel i9-9980XE, 3.0 GHz). For time steps of 0.06 s and 0.1 s, the corresponding computational times were approximately 7 min and 4 min, respectively, demonstrating favorable computational performance for long-duration time-domain simulations in irregular seas.

Table 3 Computational efficiency of OSV motion prediction in long-crested irregular waves

| Δt (s) | Physical simulation time (s) | CPU time (min) |
|---|---|---|
| 0.06 | 1800 | 7 |
| 0.1 | 1800 | 4 |

In conventional irregular-wave simulations based on regular-wave superposition, the computational cost generally increases in proportion to the number of discretized wave components in the wave spectrum. For a wave spectrum discretized into approximately 200 frequency components, repeated evaluations of wave excitation forces are required for each individual component. In contrast, the present IRF-based framework directly evaluates the instantaneous wave excitation force from the irregular-wave elevation time history through convolution integration, substantially reducing the computational cost associated with long-duration irregular-wave simulations.

The computational cost is primarily associated with the convolution integration and time-domain motion solution, while repeated evaluations of wave excitation forces for individual regular-wave components are avoided.

## 4. Motion validation in short-crested irregular waves

### 4.1 Full-scale USV and sea-trial conditions

A full-scale USV is considered for the present validation study. The main dimensions of the USV are shown in Table 4. Full-scale measurements obtained during sea trials are used to validate the present framework under realistic short-crested irregular-wave conditions.

Table 4 Particulars of the USV

| Particular | | Unit | Full scale |
|---|---|---|---|
| Length between perpendiculars | $L_{PP}$ | m | 7 |
| Breadth | $B$ | m | 2.8 |
| Draught | $\overline{d}$ | m | 0.7 |
| Height of center of gravity | $Z_g$ | m | 0.7 |
| Longitudinal center of gravity | $X_g$ | m | 3.5 |
| Mass | $m$ | kg | 7318.05 |
| Roll moment of inertia | $I_{xx}$ | $kg{\cdot}m^2$ | 6247.98 |
| Pitch moment of inertia | $I_{yy}$ | $kg{\cdot}m^2$ | 24240.31 |

The full-scale sea-trial conditions are summarized in Table 5. In these conditions, the mean heading angle represents the sailing direction of the USV, while the dominant wave direction is measured by the wave-monitoring buoy deployed in the test area. The encountered wave direction is defined as the difference between the mean heading angle of the USV and the dominant wave direction, where 180 deg and 0 deg correspond to head and following waves, respectively.

A total of 11 test conditions were recorded, all corresponding to oblique wave encounters. The maximum significant wave height and zero-crossing period during the sea trials were 0.64 m and 4.6 s, respectively, corresponding approximately to Sea State 3 conditions.

Table 5 Sea-trial conditions for the USV

| Case | Mean heading angle (deg) | Main wave direction (deg) | Encountered wave direction β (deg) | Significant wave height $H_s$ (m) | Significant wave period $T_s$ (s) | Velocity (m/s) | Froude |
|---|---|---|---|---|---|---|---|
| 1 | -3.2 | -109.0 | -105.83 | 0.53 | 4.3 | 1.06 | 0.128 |
| 2 | 1.2 | -128.0 | -129.18 | 0.61 | 4 | 1.43 | 0.173 |
| 3 | 9.5 | -128.0 | -137.51 | 0.61 | 4 | 2.16 | 0.261 |
| 4 | 11.5 | -128.0 | -139.49 | 0.61 | 4 | 2.67 | 0.322 |
| 5 | 6.4 | -128.0 | -134.41 | 0.61 | 4 | 2.90 | 0.350 |
| 6 | -1.6 | 134.3 | 135.95 | 0.64 | 4.4 | 4.51 | 0.544 |
| 7 | -5.6 | 134.3 | 139.90 | 0.64 | 4.4 | 3.71 | 0.448 |
| 8 | -7.4 | 134.3 | 141.66 | 0.64 | 4.4 | 3.25 | 0.392 |
| 9 | -8.1 | 138.2 | 146.30 | 0.6 | 4.6 | 3.00 | 0.362 |
| 10 | -7.3 | 138.2 | 145.46 | 0.6 | 4.6 | 2.58 | 0.311 |
| 11 | -13.9 | 138.2 | 152.15 | 0.6 | 4.6 | 1.74 | 0.210 |

### 4.2 Attitude-dependent hydrodynamic updating

Under oblique-wave conditions, the forward speed of the USV ranged from 1.06 m/s (corresponding to a Froude number of 0.128) to 4.51 m/s (corresponding to a Froude number of 0.544). The mean trim angle varies significantly with forward speed. Fig.7 shows the relationship between the mean trim angle and the Froude number, indicating a nonlinear increase in trim angle as the Froude number increases. When the Froude number reaches 0.544, the mean trim angle increases to 7.45 deg. The relatively large trim variation at higher speeds may alter the wetted hull geometry and hydrodynamic characteristics of the USV. Therefore, variations in the running attitude should be considered in the numerical simulations.

Based on the forward speed of the USV, the corresponding trim angle is first determined. The NURBS surface is then transformed according to the trim angle. Subsequently, an automatic station-section slicing procedure is applied to obtain uniformly distributed discrete points on each station section under different motion attitudes [30], thereby accounting for the influence of running attitude on the radiation and diffraction potentials of the USV.

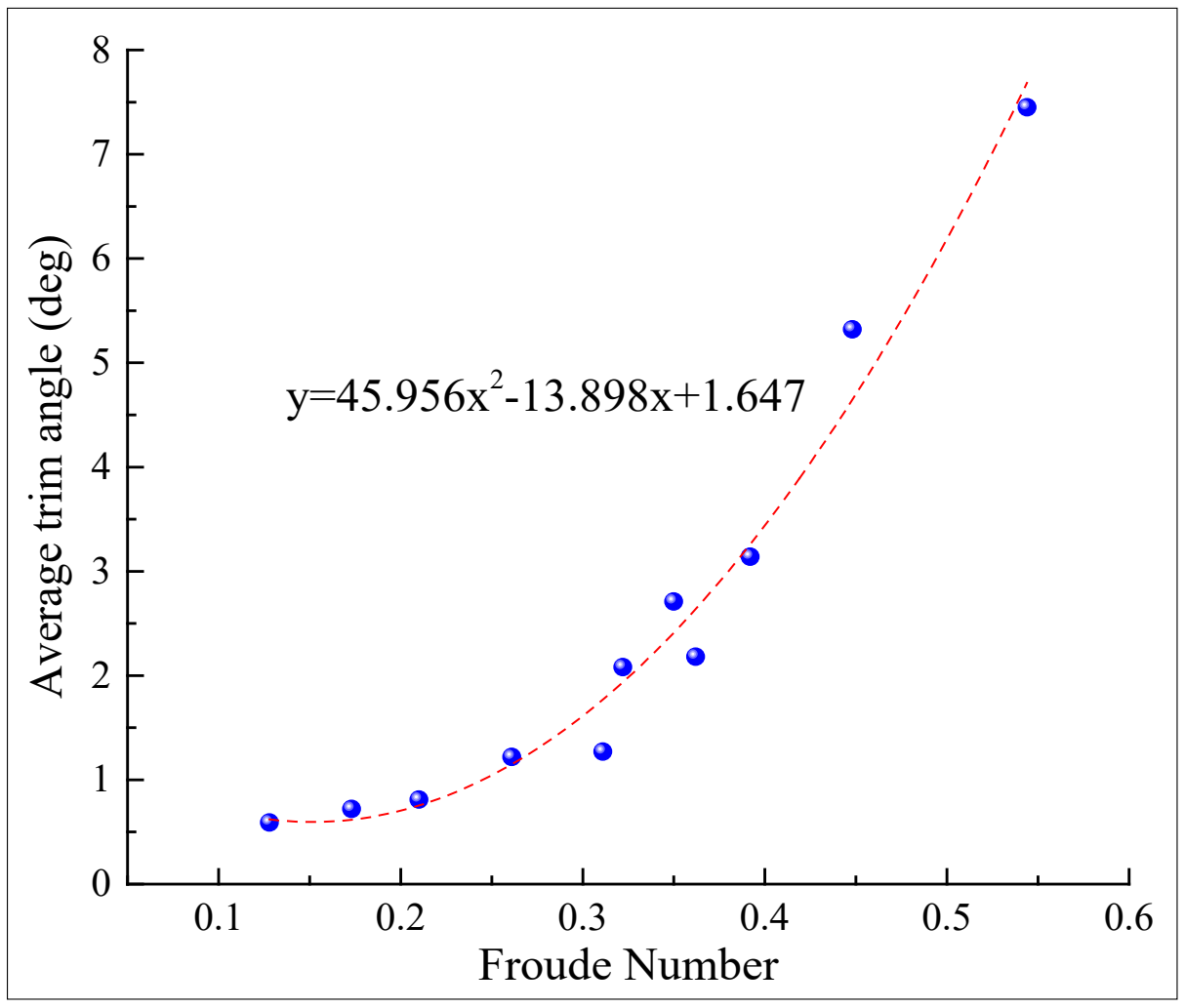

Figure 7 Relationship between Froude number and mean trim angle of the USV

### 4.3 Verification of motions in short-crested irregular waves

Since the instantaneous wave elevation encountered by the USV during sea trials cannot be directly measured, the irregular-wave elevation time histories used in the numerical simulations are reconstructed from the measured wave spectrum and corresponding statistical sea-state parameters. The reconstructed instantaneous wave elevation is then directly used as the input wave elevation in Eq. (6).

The numerical simulations are conducted using a full-scale model of the USV with a time step of 0.1 s, consistent with the sampling frequency of the full-scale measurements. For all cases, the total simulation duration is set to 200 s. The directional components of the long-crested irregular waves are discretized with an interval of 15 deg.

Figs. 8–10 compare the numerically simulated and full-scale measured time histories of the heave, roll, and pitch motions of the USV under representative operating conditions. Specifically, the heave motions are compared for Cases 5, 7, and 9; the roll motions for Cases 1, 4, and 10; and the pitch motions for Cases 2, 3, and 8. Overall, the numerical simulations reasonably reproduce the primary motion characteristics observed in the full-scale measurements. Since the reconstructed irregular-wave elevations cannot exactly reproduce the instantaneous waves encountered during the sea trials, the simulated and measured motion time histories do not coincide point-by-point. Therefore, the validation is primarily assessed from a statistical perspective, focusing on the dominant response amplitudes, characteristic periods, and overall variation trends under different operating conditions.

(a) Case5: $H_s$=0.61m $T_s$=4s $Fr$=0.35 β=-134.41deg

(b) Case7: $H_s$=0.64m $T_s$=4.4s $Fr$=0.448 β=139.9deg

(c) Case9: $H_s$=0.6m $T_s$=4.6s $Fr$=0.362 β=146.3deg

Figure 8 Comparison of numerically simulated and full-scale measured time histories of USV heave motion

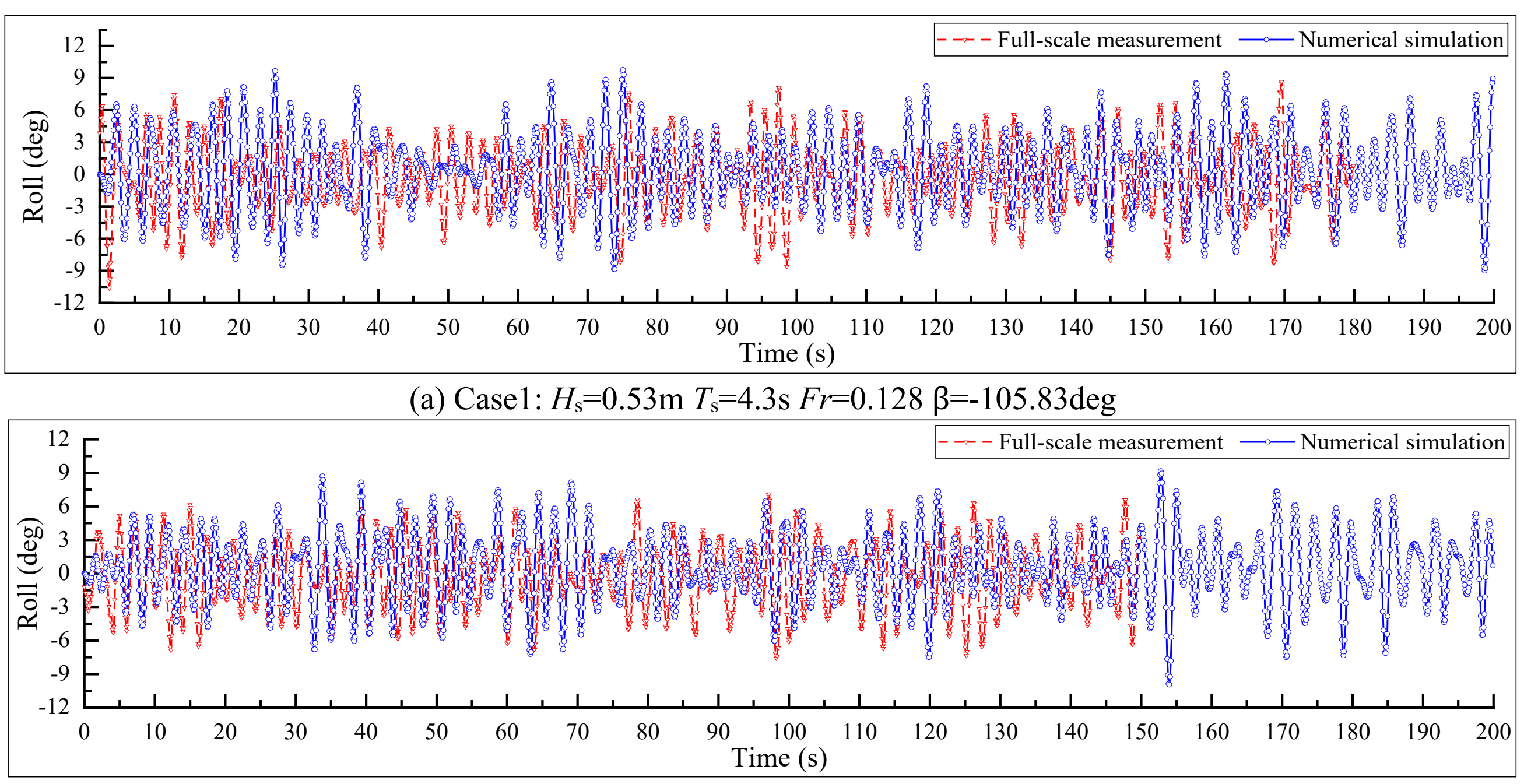

(a) Case1: $H_s$=0.53m $T_s$=4.3s $Fr$=0.128 β=-105.83deg

(b) Case4: $H_s$=0.61m $T_s$=4s $Fr$=0.322 β=-139.49deg

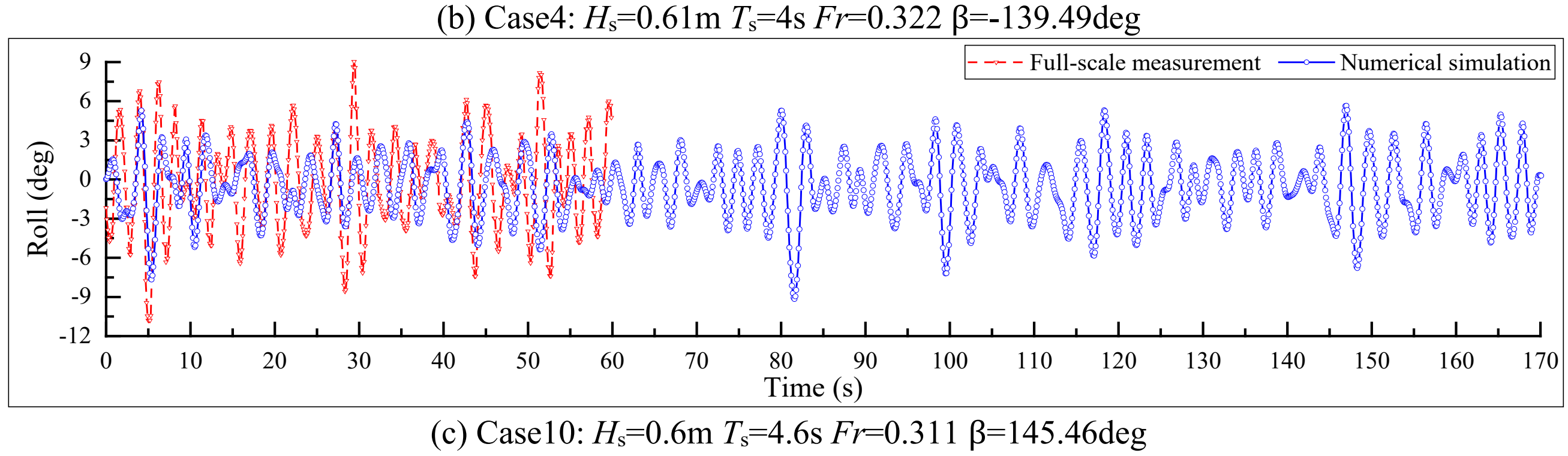

(c) Case10: $H_s$=0.6m $T_s$=4.6s $Fr$=0.311 β=145.46deg

Figure 9 Comparison of numerically simulated and full-scale measured time histories of USV roll motion

(a) Case2: $H_s$=0.61m $T_s$=4s $Fr$=0.173 β=-129.18deg

(b) Case3: $H_s$=0.61m $T_s$=4s $Fr$=0.261 β=-137.51deg

(c) Case8: $H_s$=0.64m $T_s$=4.4s $Fr$=0.392 β=141.66deg

Figure 10 Comparison of numerically simulated and full-scale measured time histories of USV pitch motion

For statistical analysis of ship motions in irregular waves, a minimum of 100 independent motion cycles is generally required to obtain statistically reliable estimates. Due to the limited record durations of 60 s, 70 s, 60 s, and 50 s for Cases 6, 9, 10, and 11, respectively, only approximately 25–35 motion cycles are available for statistical analysis in these cases. Therefore, these cases are excluded from the statistical comparison between the numerical simulations and full-scale measurements.

### 4.3.1 Significant motions

To quantitatively evaluate the statistical characteristics of the simulated motion responses, the significant values of the heave, roll, and pitch motions obtained from the numerical simulations are compared with the corresponding full-scale measurements.

Fig. 11 compares the significant values of the heave (a), roll (b), and pitch (c) motions obtained from the numerical simulations and full-scale measurements. For the heave motion, the numerical predictions show good agreement with the full-scale measurements, with most prediction errors remaining within 15%. The maximum deviation occurs in Case 8 at $Fr$ = 0.392, where the error reaches 20.60%. In the numerical simulations, when the Froude number is smaller than 0.26 (corresponding to encounter wave angles ranging from −105.8 deg to −137.51 deg), the predicted heave significant value shows a slight increasing trend with forward speed under the corresponding encounter-wave conditions. Although this trend is less evident in the full-scale measurements due to the influence of varying sea states and random wave realizations, the overall variation range and response levels are reasonably reproduced by the numerical simulations. Since the heave motion is directly governed by the vertical wave excitation, the predicted heave response is also influenced by the incident wave height. For example, when the significant wave height is relatively small ($H_s$=0.53 m, Case 1), the predicted heave significant value is the smallest (0.278 m), whereas the maximum predicted value of 0.475 m occurs in Case 8 with $H_s$=0.64 m. This variation is physically consistent with the increase in wave excitation associated with larger incident wave heights.

For the pitch motion, the numerical predictions also show good agreement with the full-scale measurements, and the overall variation trend is reasonably reproduced. The maximum deviation occurs at $Fr$=0.448, where the prediction error reaches 25.08%, while the errors for the remaining cases remain within 10%. At relatively low Froude numbers ($Fr$ <0.25), the predicted pitch motions are slightly smaller than the corresponding full-scale measurements, whereas mild overprediction is observed when $Fr$ exceeds 0.38. Since the pitch response is strongly associated with the coupled heave–pitch hydrodynamic behavior and longitudinal wave excitation moment, the predicted motion is sensitive to both the encounter-wave conditions and the variation of the running attitude. As the forward speed increases, the trim angle of the USV becomes larger, leading to changes in the wetted hull geometry and longitudinal hydrodynamic pressure distribution, which may further influence the pitch response. Overall, the numerical simulations reasonably reproduce both the response level and variation trend of the pitch motions under different operating conditions.

For the roll motion, the comparison between the numerical simulations and the full-scale measurements shows relatively larger deviations than those observed for the vertical motions. The prediction errors are generally larger than those observed for the vertical motions, with four out of seven cases remaining within 15% and the maximum deviation reaching 35.43%. According to the full-scale measurements, when the Froude number ranges from 0.13 to 0.26, the roll significant value decreases with increasing forward speed, whereas it increases again as the speed continues to rise. In contrast, the numerical simulations show an overall decreasing trend of roll motion with increasing forward speed. This trend is consistent with the commonly observed behavior of high-speed vessels, where increasing forward speed enhances viscous roll damping and hydrodynamic stabilization, thereby reducing the roll response.

The discrepancy between the numerical and measured trends may be related to the simplified treatment of viscous roll damping in the present model, as well as the influence of trim variation on the hydrodynamic characteristics of the hull. As the forward speed increases, the USV experiences a noticeable increase in trim angle, resulting in deeper stern immersion and consequently modified viscous damping characteristics. In addition, transient wave groups and variations in encounter-wave conditions during the sea trials may further contribute to the increase in measured roll amplitudes at higher forward speeds.

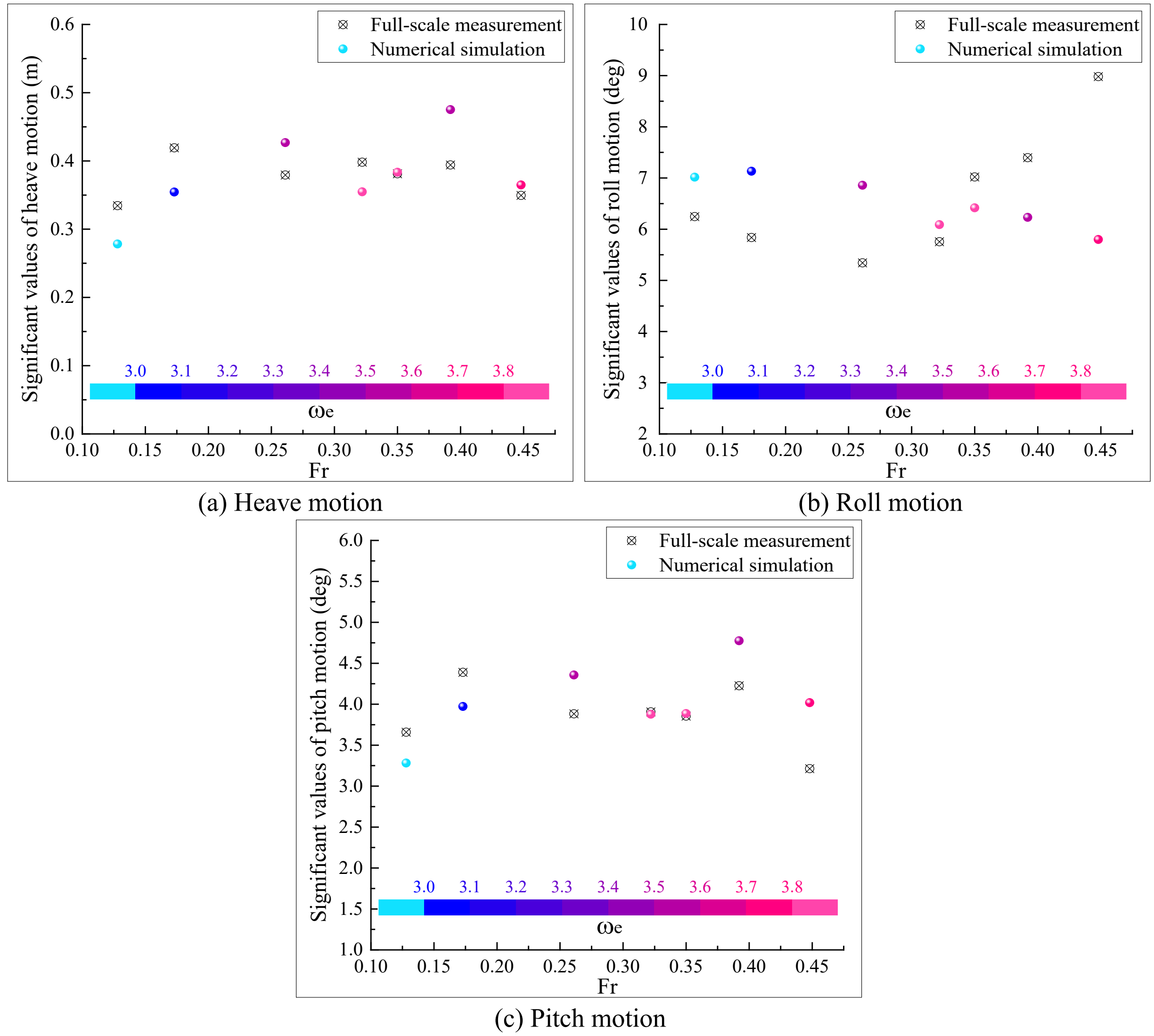

(c) Pitch motion

Figure 11 Comparison of the significant values of the heave (a), roll (b), and pitch (c) motions of the USV obtained from numerical simulations and full-scale measurements

### 4.3.2 Mean zero-crossing periods

Fig. 12 compares the mean zero-crossing periods of the heave (a), roll (b), and pitch (c) motions obtained from the numerical simulations and full-scale measurements. For the heave and pitch motions, the maximum prediction errors of the mean zero-crossing periods are −15.33% and −13.69%, respectively, both occurring in Case 4. Except for Case 4, the prediction errors of the heave mean zero-crossing periods remain within 10%, indicating that the present numerical model can reasonably capture the dominant oscillation characteristics of the vertical motions. For the roll motion, the predicted mean zero-crossing periods show good agreement with the full-scale measurements. In most cases, the prediction errors remain within 5%, with the maximum deviation reaching 14.78%. Overall, the good agreement of the mean zero-crossing periods indicates that the present numerical approach can reasonably reproduce the dominant encounter-wave frequencies and the associated hydrodynamic restoring characteristics governing the USV motion responses in irregular waves.

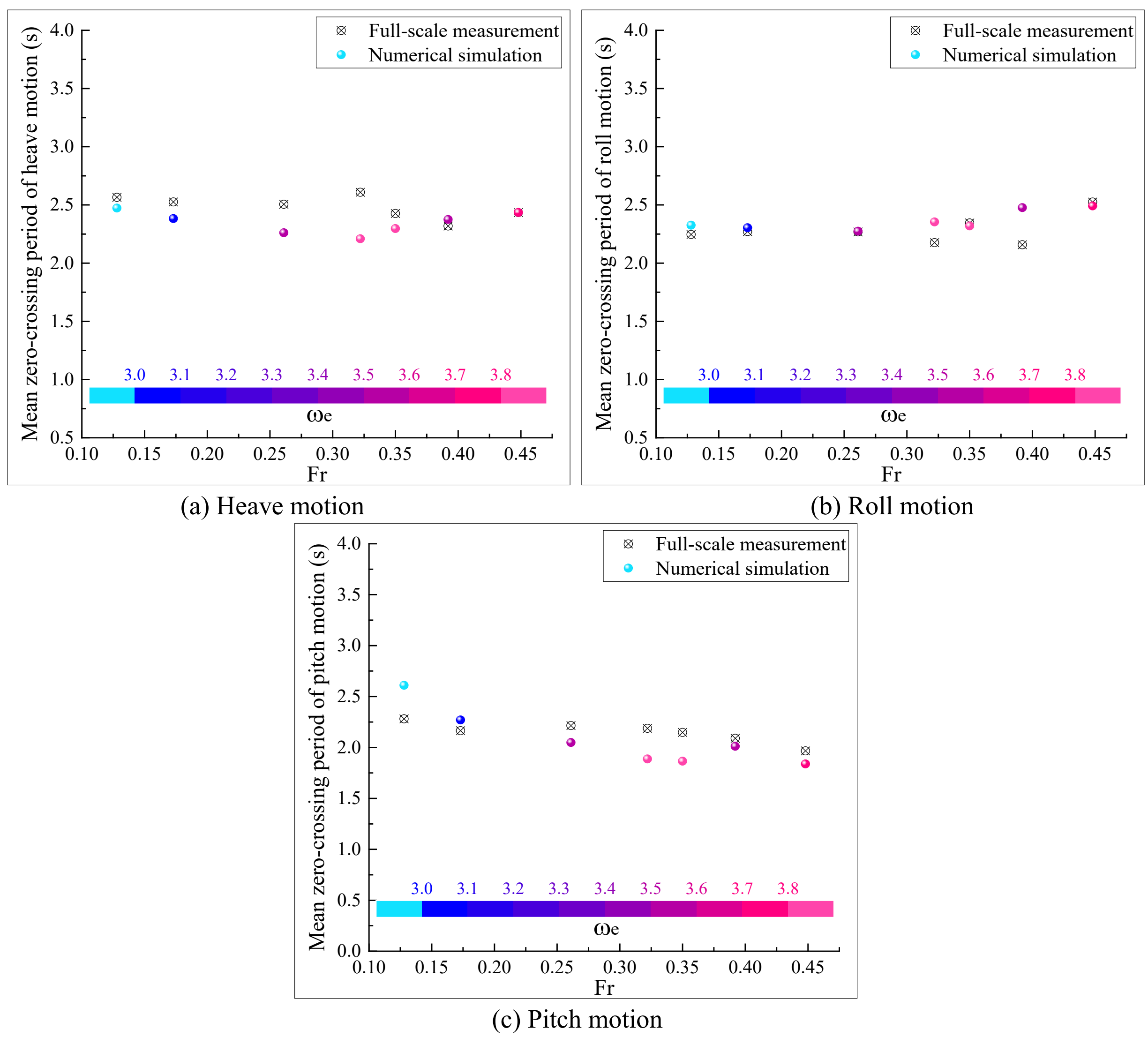

(a) Heave motion

(b) Roll motion

(c) Pitch motion

Figure 12 Comparison of mean zero-crossing period of the heave (a), roll (b), and pitch (c) motions of the USV obtained from numerical simulations and full-scale measurements

#### 4.3.3 Motion standard deviations

Table 6 presents the comparison of the motion standard deviations obtained from the numerical simulations and the full-scale measurements for the heave, roll, and pitch motions. Overall, the numerical simulations reasonably reproduce the fluctuation characteristics of the USV motions under different operating conditions, although the agreement varies among different motion modes.

For the heave motion, the predicted standard deviations generally show good agreement with the measurements. In the low-Froude-number cases (Cases 1–2), the numerical results are slightly smaller than the measured values, with errors of −13.52% and −18.58%, respectively. As the forward speed increases, the agreement improves in Cases 3–5 and 7, where the errors remain within approximately ±5%. However, in Case 8, the numerical prediction becomes noticeably larger than the measurement, with an error of 29.73%. Since the heave fluctuation is directly governed by the vertical wave excitation, relatively small differences in the reconstructed irregular-wave elevation and encounter-wave conditions may lead to larger deviations in the predicted fluctuation amplitude.

For the roll motion, relatively larger deviations are observed between the numerical simulations and the full-scale measurements. In the lower-speed cases (Cases 1–3), the predicted roll standard deviations are consistently larger than the measured values, with errors ranging from 13.28% to 27.37%. As the forward speed increases, the discrepancy gradually decreases, and the numerical predictions become closer to the measurements in Cases 4 and 5. In the higher-speed cases (Cases 7–8), the numerical simulations tend to underpredict the roll fluctuations, with errors of −35.46% and −17.10%, respectively. Similar to the significant roll motions discussed in Section 4.3.1, these discrepancies are mainly associated with the simplified treatment of viscous roll damping and the influence of running-attitude variation on the hydrodynamic damping characteristics.

For the pitch motion, the agreement between the numerical simulations and the full-scale measurements is generally better. Most prediction errors remain within ±10%, indicating that the present numerical model can reasonably reproduce the fluctuation characteristics of the pitch response. Since the pitch motion is primarily governed by coupled heave–pitch restoring and excitation effects, which are reasonably represented in the present potential-flow-based framework, the predicted pitch fluctuations remain comparatively stable under different operating conditions.

Table 6 Comparison of motion standard deviations obtained from numerical simulations and full-scale measurements

| Case | Motion | $\sigma = \sqrt{\frac{1}{N-1}\sum_{i=1}^{N}(x_i - \bar{x})^2}$ | | Error |
|---|---|---|---|---|
| | | full-scale measurement | numerical simulation | |
| 1 | heave | 0.160 | 0.138 | -13.52% |
| | roll | 3.195 | 3.619 | 13.28% |
| | pitch | 1.800 | 1.629 | -9.50% |
| 2 | heave | 0.205 | 0.167 | -18.58% |
| | roll | 2.934 | 3.517 | 19.90% |
| | pitch | 2.203 | 1.939 | -11.96% |
| 3 | heave | 0.194 | 0.208 | 7.58% |
| | roll | 2.679 | 3.412 | 27.37% |
| | pitch | 1.998 | 2.080 | 4.12% |
| 4 | heave | 0.180 | 0.177 | -1.64% |
| | roll | 2.938 | 3.150 | 7.18% |
| | pitch | 1.918 | 2.033 | 6.03% |
| 5 | heave | 0.185 | 0.190 | 2.82% |
| | roll | 3.425 | 3.312 | -3.32% |
| | pitch | 1.973 | 1.955 | -0.94% |
| 7 | heave | 0.184 | 0.191 | 4.06% |
|   | roll | 4.435 | 2.862 | -35.46% |
|   | pitch | 1.746 | 2.009 | 15.06% |
|   | heave | 0.182 | 0.236 | 29.73% |
| 8 | roll | 3.848 | 3.190 | -17.10% |
|   | pitch | 2.148 | 2.274 | 5.88% |

Overall, the comparisons of the significant motion amplitudes, mean zero-crossing periods, and motion standard deviations demonstrate that the present numerical framework can reasonably reproduce the primary statistical characteristics of USV motions in short-crested irregular waves. Good agreement is achieved for the heave and pitch motions in both response amplitudes and characteristic periods, while the larger discrepancies observed in the roll motion are mainly associated with the simplified treatment of viscous roll damping and the attitude-dependent hydrodynamic effects at different forward speeds. Despite these discrepancies, the numerical results generally capture the dominant variation trends of the measured motion responses under different sea states and operating conditions. Owing to the efficient IRF-based evaluation of wave excitation forces, the present framework provides a computationally efficient approach for predicting transient USV motions in realistic irregular-wave environments. In the present framework, the computational cost mainly depends on the number of discretized directional components rather than the number of frequency components in the wave spectrum, which substantially improves the efficiency of directional irregular-wave simulations. The achieved efficiency makes the method suitable for high-fidelity simulation environments and potential real-time motion prediction applications.

### 4.4 Discussion on method applicability and limitations

This study presents a computationally efficient framework for predicting instantaneous USV motions in short-crested irregular waves and validates the method against full-scale measurements. Overall, the framework provides reasonable motion predictions under oblique short-crested wave conditions, while several limitations should still be considered in practical engineering applications.

(1) Theoretical limitations of the potential-flow framework

The present method is primarily based on potential-flow theory and therefore relies on the assumption of weakly nonlinear ship motions. Under conditions involving large roll motions, strongly nonlinear waves, or green-water phenomena, nonlinear effects such as flow separation, viscous vortex shedding, and strongly deformed free surfaces cannot be fully represented, which may reduce prediction accuracy. In addition, at relatively high forward speeds, the influence of speed-dependent hydrodynamic effects, including lift-induced forces and nonlinear pressure redistribution around the hull, becomes increasingly significant and may further increase the discrepancy between numerical predictions and measured responses.

(2) Uncertainty associated with damping and environmental conditions

To compensate for the limitations of the potential-flow formulation, viscous damping effects are incorporated through additional damping models. However, the damping coefficients remain partially empirical and may vary under different forward speeds and motion conditions. In particular, the coupled effects of trim variation, flow separation near the stern, and wake-vortex evolution may introduce additional uncertainty into the roll-motion prediction. Moreover, the environmental conditions encountered during full-scale sea trials contain inherent randomness. Variations in local wind fields, wave grouping, and encounter-wave conditions may influence the measured vessel responses and contribute to deviations between numerical simulations and full-scale measurements. Therefore, part of the observed discrepancies may originate from environmental uncertainty rather than from the numerical framework itself.

(3) Practical applicability of the present method

Despite these limitations, the present framework provides a physically consistent and computationally efficient approach for predicting transient USV motions in realistic irregular-wave environments. The method is particularly suitable for long-duration time-domain simulations, parametric studies, preliminary design assessment, and motion-trend evaluation under moderate sea conditions. However, caution should be exercised when applying the method to strongly nonlinear, high-speed, or extreme-wave conditions, where additional viscous and nonlinear hydrodynamic effects may become dominant.

## 5. Effect of directional spectrum discretization on motion prediction

To evaluate the influence of directional-spectrum representation on motion prediction, numerical simulations were performed using different directional discretization intervals. In practical simulations of short-crested irregular waves, the continuous directional spectrum is approximated by a finite number of directional wave components, and the corresponding directional resolution may affect the reconstructed wave field and the predicted motion responses.

To ensure a consistent comparison, the instantaneous wave amplitudes of the long-crested irregular-wave components were kept identical under each discretization condition so that the influence of directional discretization could be isolated. Fig. 13 presents the comparison of the significant values of the heave, roll, and pitch motions of the USV under short-crested irregular wave conditions with directional discretization intervals of 15 deg, 30 deg, and 45 deg. From the comparison of the significant values of the heave and pitch motions, it can be observed that, for most cases, the predictions obtained with a directional discretization interval of 30 deg remain in good agreement with the full-scale measurements for most cases.

For the roll motion, the significant values predicted using a 30 deg discretization interval differ by less than 10% compared with those obtained using a 15 deg interval, indicating that further refinement of the directional resolution has a limited impact on the roll response prediction. Similarly, the differences in the predicted heave and pitch significant values between the 15 deg and 30 deg discretizations are generally small, with a maximum deviation of approximately 14%. These results suggest that the motion responses are not highly sensitive to the directional discretization once a certain resolution threshold is reached. However, when the directional discretization interval is increased to 45 deg, noticeable discrepancies begin to appear. For example, in Case 5 ($Fr$=0.35), the predicted significant values of the heave and roll motions differ by 19% and 17%, respectively, compared with those obtained using a 15 deg discretization. This can be attributed to the insufficient representation of the directional spreading of wave energy at coarse resolution. With larger discretization intervals, the wave energy is concentrated into fewer directional components, which reduces the accuracy of the reconstructed wave field and may lead to distortions in the phase relationship and superposition of wave components. As a result, the predicted wave excitation forces and the corresponding motion responses deviate more significantly from the reference results.

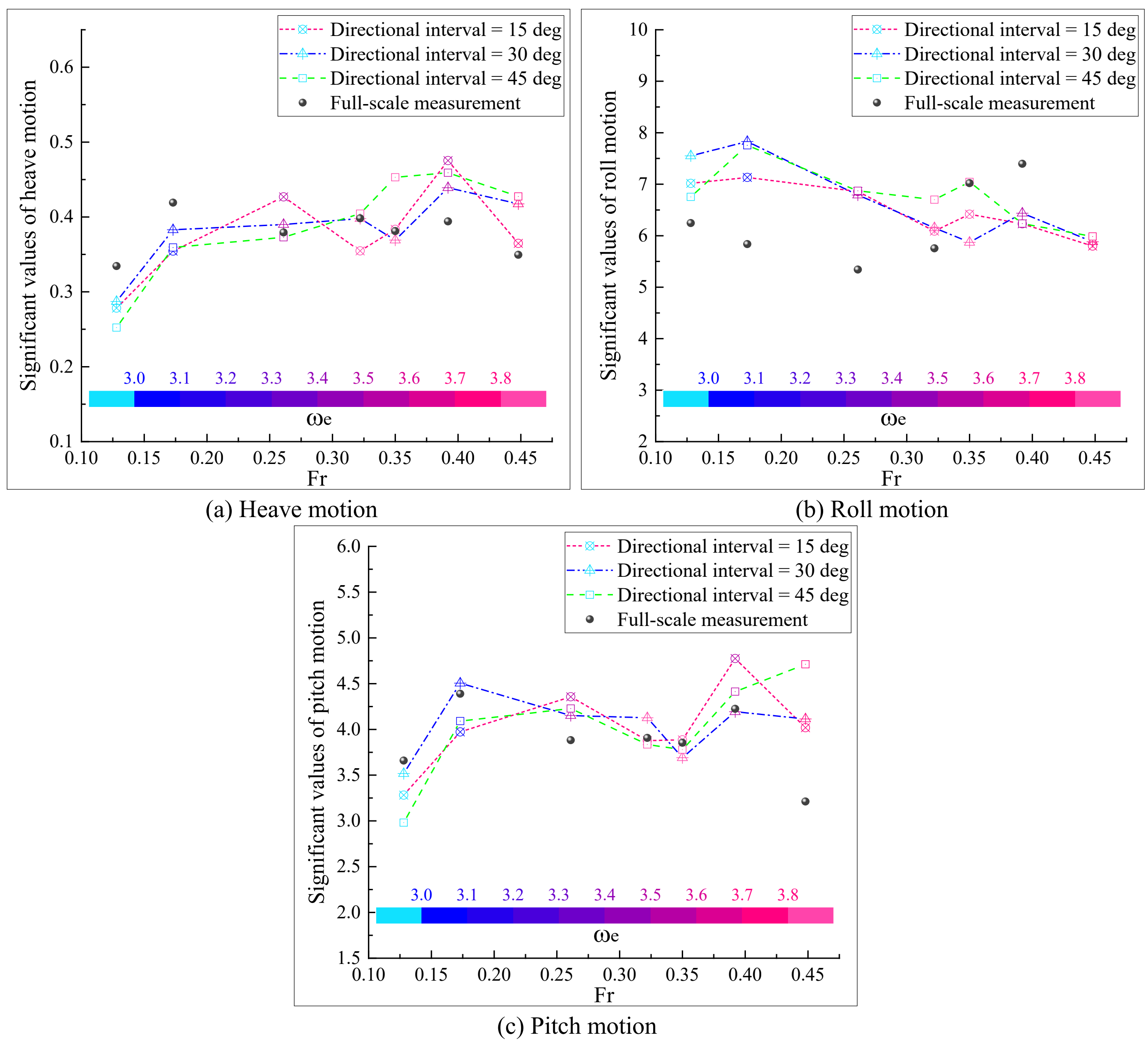

(a) Heave motion (b) Roll motion

(c) Pitch motion

Figure 13 Comparison of directional discretization interval on the significant motions of the USV under short-crested irregular waves: (a) heave, (b) roll, and (c) pitch

Figure 14 compares the mean zero-crossing periods of the heave, roll, and pitch motions of the USV obtained under different directional discretization intervals. It can be observed that the influence of the directional discretization interval on the mean zero-crossing periods is generally limited. For most cases, the differences in the predicted mean zero-crossing periods between the 45 deg and 15 deg discretization intervals remain within 5% for all three motion modes.

For the roll motion, the mean zero-crossing periods predicted using the 15 deg discretization interval show the best agreement with the full-scale measurements. However, the differences between the results obtained using 30 deg and 15 deg discretizations are also within 5% for most cases, indicating that further refinement of the directional resolution has only a minor impact on the predicted motion periods.

From a physical perspective, the mean zero-crossing period of ship motions is primarily governed by the encounter wave frequency and the inherent dynamic characteristics of the vessel, such as hydrostatic restoring forces and inertia properties. Compared with motion amplitudes, the motion periods are less sensitive to the detailed directional distribution of wave energy. Therefore, even with a relatively coarse directional discretization, the dominant frequency content of the wave field can still be adequately captured, leading to only minor variations in the predicted motion periods.

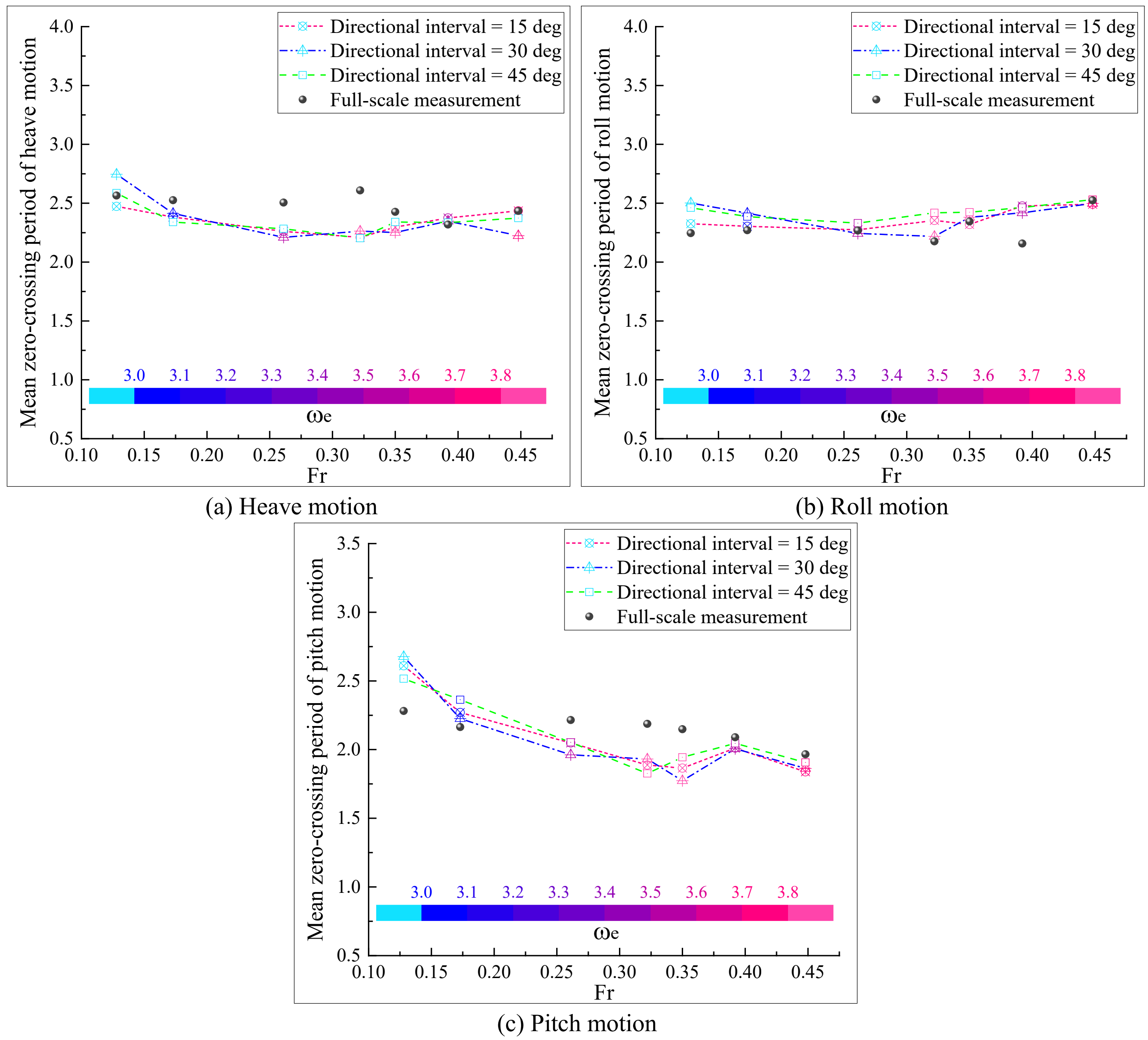

Figure 14 Comparison of directional discretization interval on mean zero-crossing period of the USV under short-crested irregular waves: (a) heave, (b) roll, and (c) pitch

Figure 15 and Figure 16 compare the motion time histories obtained using directional discretization intervals of 15 deg and 30 deg under the lowest-speed ($Fr$=0.128, Case 1) and highest-speed ($Fr$=0.448, Case 7) conditions, respectively. Overall, the two discretization intervals produce similar motion frequencies and comparable phase variations over most of the simulation duration, indicating that the dominant dynamic characteristics are reasonably captured even with the coarser directional resolution. Nevertheless, local discrepancies in the motion amplitudes and peak responses can still be observed in certain time intervals. Compared with the 15 deg discretization, the 30 deg discretization tends to produce slightly larger local peak responses. This behavior is attributed to the reduced resolution in representing the directional spreading of wave energy. With fewer directional components, the reconstructed wave field becomes less smooth, which may enhance local constructive interference among wave components. In contrast, a finer discretization distributes wave energy more evenly across directions, resulting in a more diffused superposition and smoother extreme responses. This indicates that directional discretization primarily affects the representation of extreme events through wave energy redistribution, while having limited influence on the dominant motion frequencies.

Considering that the 30 deg discretization yields motion predictions close to those obtained with finer directional resolution while significantly reducing the computational cost, it is regarded as a reasonable compromise between computational efficiency and prediction accuracy for practical USV motion simulations in short-crested irregular waves.

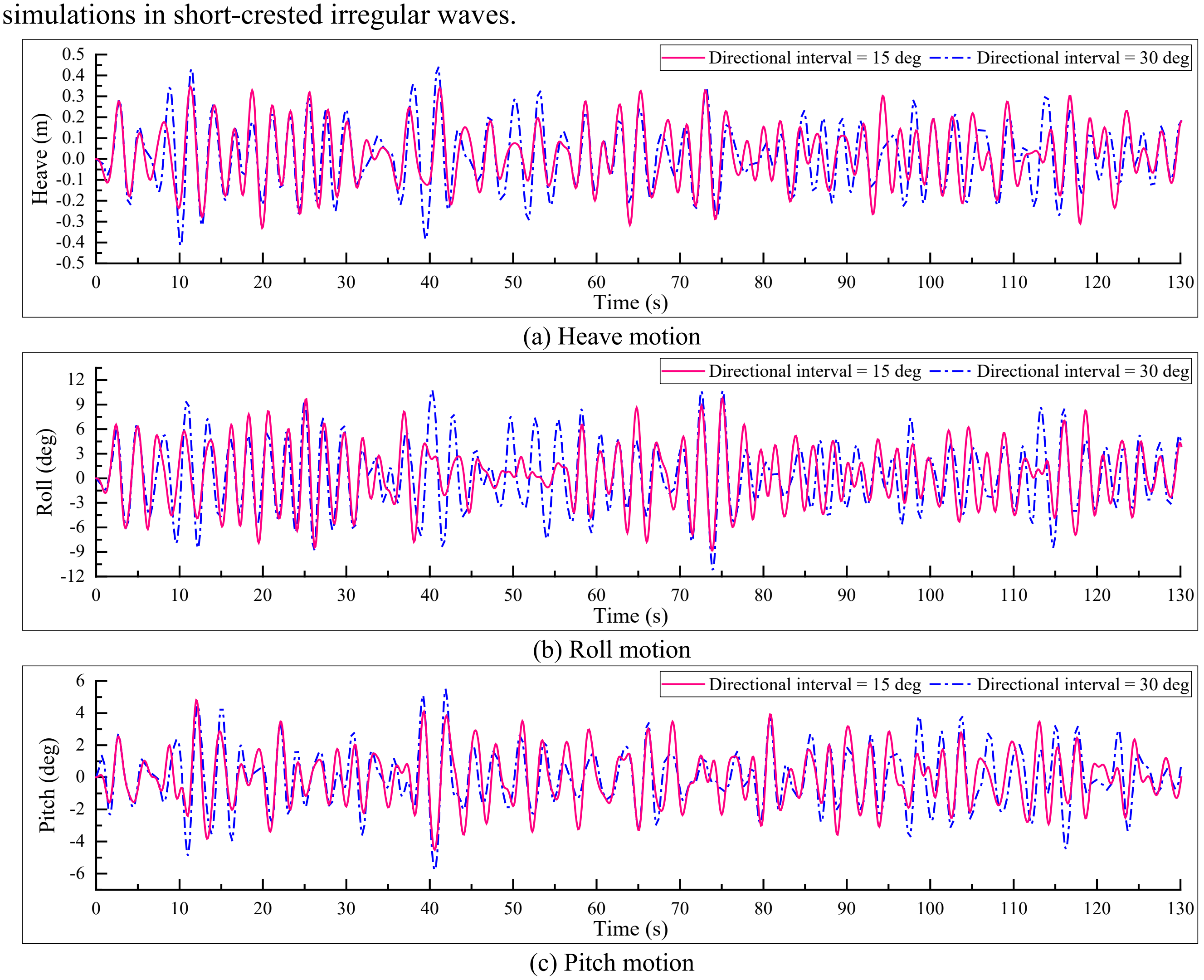

Figure 15 Comparison of the time histories of USV motions using directional discretization intervals of 15 deg and 30 deg (Case1: $H_s$=0.53m $T_s$=4.3s $Fr$=0.128 β=-105.83deg)

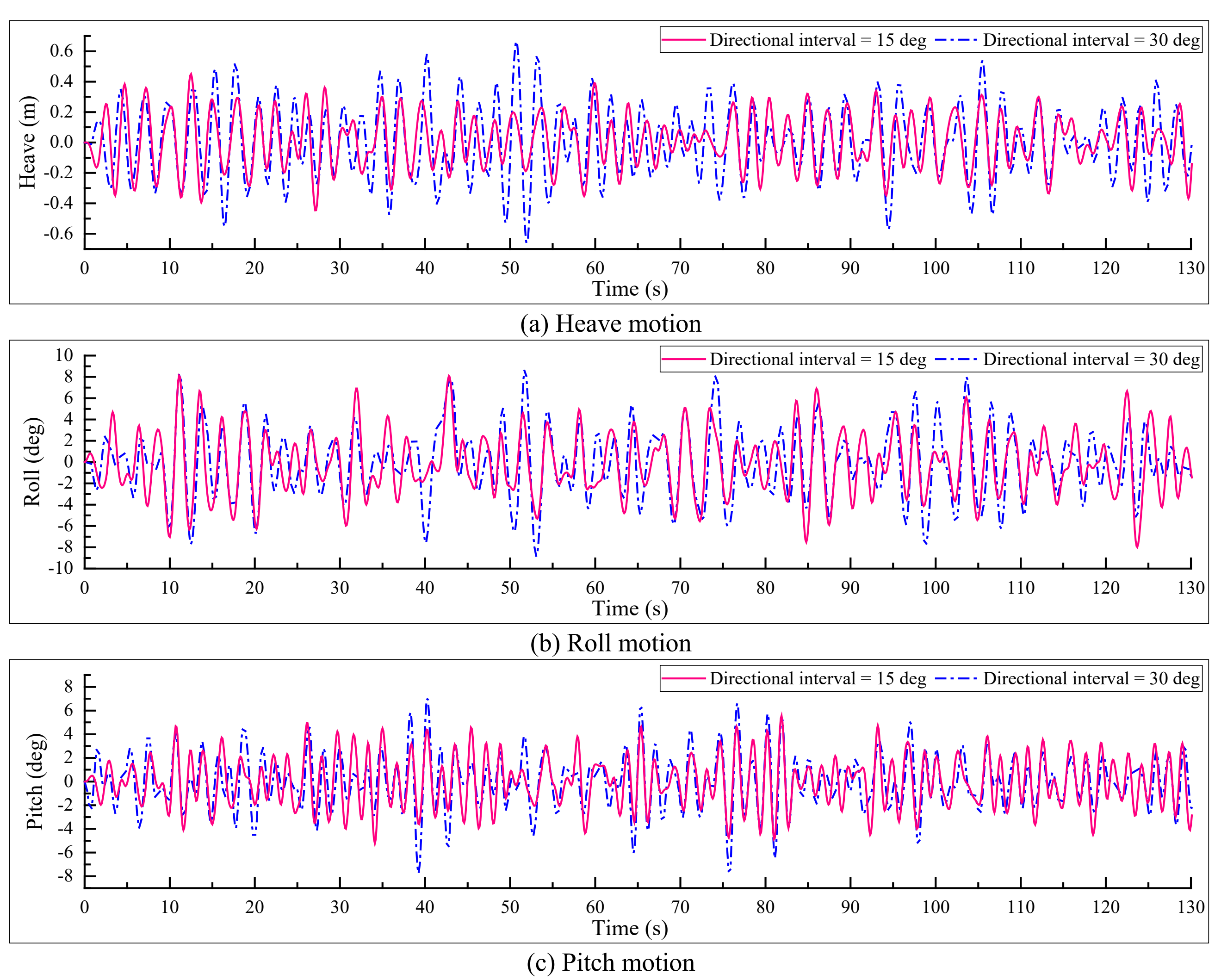

(a) Heave motion

(b) Roll motion

(c) Pitch motion

Figure 16 Comparison of the time histories of USV motions using directional discretization intervals of 15 deg and 30 deg (Case7: $H_s$=0.64m $T_s$=4.4s $Fr$=0.448 β=139.9deg)

## 6. Conclusions

This study presents a computationally efficient time-domain framework for predicting USV motions in short-crested irregular waves. By transforming frequency-domain hydrodynamic coefficients into time-domain convolution kernels and incorporating instantaneous nonlinear restoring forces through wetted-surface pressure integration, the present framework captures weakly nonlinear motion characteristics while maintaining physical consistency.

Compared with conventional irregular-wave simulations based on regular-wave superposition, the present framework directly evaluates the instantaneous wave excitation force in the time domain through convolution integration. As a result, the computational cost mainly depends on the number of discretized directional components rather than the number of frequency components in the wave spectrum, substantially improving the computational efficiency for long-duration simulations in directional irregular waves.

Validation against both model-test and full-scale measurement data demonstrates that the present framework can reasonably reproduce the primary statistical characteristics of ship motions in irregular waves. For the OSV motions in long-crested irregular waves, the present IRF-based framework accurately captures the transient roll responses while maintaining high computational efficiency. For the full-scale USV validation in short-crested irregular waves, good agreement is achieved for the heave and pitch responses, while the larger discrepancies observed in roll motion are mainly associated with viscous damping uncertainty and attitude-dependent hydrodynamic effects at different forward speeds.

In addition, the sensitivity analysis of directional-spectrum discretization indicates that motion amplitudes exhibit moderate dependence on directional resolution, whereas the motion periods are relatively insensitive to the discretization interval. A directional discretization interval of 30 deg is found to provide a reasonable compromise between computational efficiency and prediction accuracy for practical USV motion simulations in short-crested irregular waves.

Overall, the present framework provides a physically consistent and computationally efficient approach for transient motion prediction of USVs operating in realistic irregular-wave environments. The method is suitable for long-duration time-domain simulations, parametric studies, preliminary design assessment, and potential real-time prediction applications under moderate sea conditions. However, caution should be exercised when applying the framework to strongly nonlinear, high-speed, or extreme-wave conditions, where additional viscous and nonlinear hydrodynamic effects may become significant.

## Acknowledgements

The authors sincerely appreciate the contributions of all staffs and students involved in the full-scale USV experiments. Authors are grateful to the staff at the Multiple Function Towing Tank of Shanghai Jiao Tong University and students participated in this test. This research is supported by the Postdoctoral Fellowship Program of CPSF under Grant Number GZC20240983, the China Postdoctoral Science Foundation under Grant Number 2025M771840.